\documentstyle[floats,epsfig,aps,prb]{revtex}

\newcommand{\dfrac}[2]{\frac{\strut \displaystyle{#1}}{\displaystyle{#2}}}

\begin{document}

\draft

%%%%%%%%%%%%%%%%
\twocolumn[\hsize\textwidth\columnwidth\hsize\csname
@twocolumnfalse\endcsname

\title{\large\bf Magnetization plateaus as insulator-superfluid
  transitions in quantum spin systems}

\author{Tsutomu Momoi\cite{address1}}
\address{Lyman Laboratory of Physics, Harvard University, Cambridge, 
MA 02138}
\author{Keisuke Totsuka\cite{address2}}
\address{ Magnetic Materials Laboratory, The Institute of Physical
 and Chemical Research (RIKEN),
 Wako, Saitama, 351-0198 Japan}
\date{\hspace*{5cm}}
%\date{\today}

\maketitle

\begin{abstract}
We study the magnetization process in two-dimensional $S=1/2$ spin
systems, to discuss the appearance of a plateau structure.
The following three cases are considered:
1) the Heisenberg antiferromagnet
and multiple-spin exchange model on the triangular lattice, 2)
Shastry-Sutherland type lattice, (which is a possible model for
SrCu$_2$(BO$_3$)$_2$,) 3) 1/5-depleted lattice (for CaV$_4$O$_9$).
We find in these systems that magnetization
plateaus can appear owing to a transition from superfluid to a Mott
insulator of magnetic excitations.
The plateau states have CDW order of the
excitations. The magnetizations of the plateaus depend on components of
the magnetic excitations, range of the repulsive interaction, and the
geometry of the lattice.
\end{abstract}
\vskip-2mm
%%%%%%%%%%%%%%%
]
\narrowtext

In some one-dimensional spin systems, spin-density-wave states
with finite spin gap appear under a finite magnetic field
accompanying plateau structures in the
magnetization process.
Magnetization plateaus were observed in some quasi
one-dimensional materials.\cite{NarumiHSKNT}
Theoretical arguments clarify that the
appearance of the plateau is explained by an
insulator-conductor transition of magnetic excitations.\cite{Totsuka}
In two- or higher-dimensional systems, magnetization plateaus have
been also found in both theoretical%
\cite{NishimoriM,Chubukov-Nikuni,KuboM}- and
experimental studies.\cite{SuematsuOSSMD,NojiriTM,Kageyama}
In this paper, we propose a rather general picture that these
two-dimensional plateaus are formed owing to field-induced
insulator-superfluid transitions of magnetic excitations.
To demonstrate how it works, we discuss three examples in details.

The first example is a family of antiferromagnets on a triangular
lattice. For the $S=1/2$ antiferromagnet on a triangular lattice
(AFT), Nishimori and Miyashita\cite{NishimoriM} found a
magnetization plateau at $m/m_{\rm sat}=1/3$,
which comes from the appearance of
a collinear state with three sublattices, i.e., the so-called ``uud''
state. This plateau was actually observed in AFT materials like
C$_6$Eu(Ref.~\onlinecite{SuematsuOSSMD}) and
CsCuCl$_3$(Ref.~\onlinecite{NojiriTM}).
Recently in a multiple-spin exchange (MSE) model, which is a possible
model\cite{MSE} for solid $^3$He films, a
magnetization plateau was predicted\cite{KuboM} at
$m/m_{\rm sat}=1/2$.
In this case, the plateau is attributed to the formation of a
similar collinear state but with four sublattices.
The magnetization processes of these systems have been studied
extensively and here we just attempt {\em interpreting}
the known results to test the new picture.

\begin{figure}[tbp]
  \begin{center}
    \leavevmode
    \epsfig{file=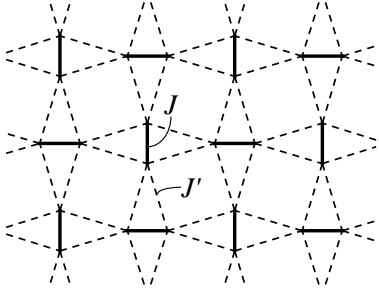,width=3.3in}
  \end{center}
  \caption{Shastry-Sutherland lattice}
  \label{fig:S-Slattice}
\end{figure}

\begin{figure}[tbp]
  \begin{center}
    \leavevmode
    \epsfig{file=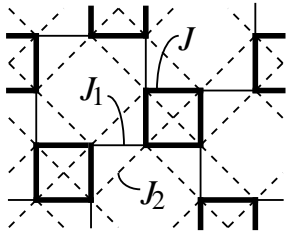,width=3.3in}
  \end{center}
  \caption{1/5-depleted square lattice.}
  \label{fig:1/5-depleted}
\end{figure}

We take as the second example
the $S=1/2$ Heisenberg antiferromagnet
(HAF) on the Shastry-Sutherland lattice (Shastry-Sutherland model,
hereafter. See Fig.~\ref{fig:S-Slattice}),\cite{ShastryS} 
which is known to have an exact dimer ground state.
Recently Kageyama et al.\cite{Kageyama} found that
SrCu$_2$(BO$_3$)$_2$ realizes a lattice structure equivalent
to that discussed in Ref.\onlinecite{ShastryS} and that it
has a gapful ground state.
The magnetization measurements
show plateaus at $m/m_{\rm sat}=1/8$ and 1/4.
The last is the $S=1/2$ HAF on the 1/5-depleted square lattice
(Fig.~\ref{fig:1/5-depleted}),
which includes a model Hamiltonian
for CaV$_4$O$_9$.
In this system, the plaquette singlet state
is realized in the ground state.\cite{SigristU}

%%%
In our picture, the plateau states can be regarded as Mott
insulators of effective magnetic particles; repulsive interactions
induce various kinds of charge-density-wave (CDW)
long-range order leading
to a finite energy gap in particle-hole excitations.
Except for the plateau phases, magnetic particles are
conducting to form supersolid, in which superfluidity and CDW
coexist, and magnetization increases smoothly.
Of course, the charge density is translated into
the spin ($S^{z}$) density and superfluidity here
means long-range order in the direction perpendicular
to the field.
Although this essential picture is common to the three examples,
the concrete forms of the magnetic particles are different.

In the first example, i.e.\ AFT and an MSE model,
a single flipped spin itself works as the magnetic particle,
while the triplets on the dimer ($J$) bonds are the relevant
particles of the Shastry-Sutherland
model.
We find plateaus at $m/m_{\rm sat}=1/2$ and 1/3.
In the third one, i.e. the 1/5-depleted square lattice,
plaquette triplets behave as particles.
We predict plateaus at $m/m_{\rm sat}=1/8$, 1/4, 1/2.

The magnetizations where a plateau appears depend on (i)
the form of the magnetic
excitations, (ii) range of the repulsion between them, and
(iii) the geometry of the lattice.
Finally we summarize common features on properties of the phase
transition.

\underline{\it Spin = magnetic particle:} The magnetization plateau
for the AFT system was found in spin 1/2
anisotropic\cite{NishimoriM} and isotropic\cite{Chubukov-Nikuni} Heisenberg
models. The Hamiltonian is
\begin{equation}\label{eq:XXZ}
H=J\sum_{\langle i,j \rangle}(S_i^x S_j^x + S_i^y S_j^y
+ \eta S_i^z S_j^z) - B\sum_i S_i^z,
\end{equation}
where the summation runs over all nearest-neighbor pairs and $B$
denotes the magnetic field.
For $\eta \ge 1$, the magnetization curve has
a plateau at
$m/m_{\rm sat}=1/3$. The ground state in the plateau phase
is of a collinear
structure with three sublattices,
where two of three spins direct upward and
the other downward.  In the other phases, magnetic states have
non-collinear structures with off-diagonal long-range order
(ODLRO).
If we introduce a particle picture, i.e., recognize spin dynamics as
induced by the motion of a certain kind of particles,\cite{MatsubaraM}
the appearance of plateau is
easily understood from simple consideration about compressibility of
the particles.
Regarding an up spin as a hard-core boson and a down one as a
vacancy,\cite{MatsubaraM} we can rewrite the Hamiltonian
(\ref{eq:XXZ}) as
\begin{equation}
H = \dfrac{J}{2} \sum_{\langle i,j \rangle} \{  (b_i^\dagger b_j + h.c.)
+ 2\eta n_i n_j \}
- (B + 3J\eta) \sum_{i} n_i,
\end{equation}
where $b_i^\dagger$ denotes the creation operator of the hard-core
boson on site $i$, and $n_i$ the number operator.
Since particles carry magnetic moment unity, the chemical
potential
$\mu=B+3J\eta$ is controlled by the magnetic field and the
$\mu$-dependence of the particle density $n$
corresponds to the magnetization curve of
the original spin system.

\begin{figure}[tbp]
  \begin{center}
    \leavevmode
    \epsfig{file=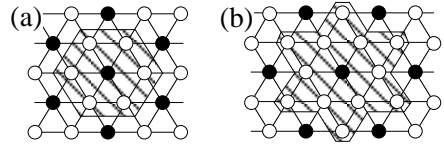,width=3.3in}
  \end{center}
  \caption{Spin-density-wave order in AFT (a) and MSE model (b). Black
    (white) circles denote down (up) spins.}
  \label{fig:CDW}
\end{figure}

The hopping term comes from the spin exchange (XY)
term and the repulsive interaction from the diagonal (Ising)
part. The anisotropic case,
$\eta>1$, is mapped to the strong coupling ({\it i.e.} strong
repulsion) region of the corresponding boson system.
The particle-hole transformation converts
the system into that of holes with repulsion of the same strength;
in the strong coupling limit, the ground state
at the filling $n=2/3$ ($m/m_{\rm sat}=1/3$) has
the density wave long-range order,
with the three-sublattice structure shown in Fig.~\ref{fig:CDW}(a).
Due to the repulsive interaction, this state is
incompressible, i.e.
$dn/d\mu = 0$, and density-fluctuation energy has a finite gap
above the ground state. Except for the filling(s) $n=2/3$ (and 1/3),
there are vacancies and hence particles are mobile (conducting).
Since the particles
obey boson statistics and the system is uniform, the system presumably
shows superfluidity. There is perfect correspondence between the
above consideration and the previous
results;\cite{NishimoriM,Chubukov-Nikuni} the
insulating CDW state with $dn/d\mu=0$, is consistent
with the spin collinear state, where susceptibility is
vanishing.
On the other hand, superfluidity of bosons corresponds
to non-collinear ODLRO of spins.

The particle density where CDW stabilizes depends on the range of repulsive
interactions.
To see this, we next discuss MSE model
with four-spin exchange on the
triangular lattice, where repulsion acts further than
in the Heisenberg model.
The Hamiltonian is given by
\begin{equation}\label{Hamiltonian}
{\cal H}= J \sum_{\langle i,j \rangle}
        \mbox{\boldmath $\sigma$}_i \cdot \mbox{\boldmath $\sigma$}_j
  + K \sum_p h_p - B\sum_i \sigma_i^z,
\end{equation}
where $\mbox{\boldmath $\sigma$}_i$ denote Pauli matrices.
The second summation runs over all minimum diamond clusters and $h_p$
is the four-spin exchange $h_p = 4(P_4 + P_4^{-1})-1$ with
the ring permutation of four spins $P_4$.
It was shown that three- and four-spin exchange
interactions are very strong in two-dimensional solid $^3$He
due to strong quantum fluctuations.\cite{MSE}
Theoretically a magnetization plateau was found\cite{KuboM} at
$m/m_{\rm sat}=1/2$ instead of $m/m_{\rm sat}=1/3$.
In the particle picture, bosons feel the following
two-body repulsion
\begin{equation}
V = 4(J+5K)\sum_{\langle i,j \rangle} n_i n_j
+ 4 K \sum_{( i,j ) \atop \in N.N.N.} n_i n_j.
\end{equation}
The repulsive interaction acts in both nearest- and
next-nearest-neighbor sites. Figure~\ref{fig:CDW}(b) shows the region
where the interaction works. Because of the range of repulsion,
the particles can solidify at the density $n=1/4$.
(Note that the solidification occurs only if the repulsion
overcomes the effect of the hopping term.)
This insulating phase at $n=1/4$ corresponds to the magnetization
plateau at $m/m_{\rm sat}=1/2$ in the original spin system.
The previous numerical result in Ref.\ \onlinecite{KuboM} on the
ground state of the plateau phase is
consistent with the CDW order shown in Fig.~\ref{fig:CDW}(b).
%It was also shown that the spin excitation spectrum in
%$S_{zz}(k,\omega)$ has a finite gap above the plateau
%state.\cite{KuboM}

\underline{\it Dimer triplet:}
%When two spins are strongly coupled with the antiferromagnetic
%interaction, the dimer singlet state realizes in the ground state.
When specific pairs of two spins are more strongly coupled than
to the others by an antiferromagnetic interaction,
the dimer singlet state realizes in the ground state.
Under a weak magnetic field, $S^z=1$ triplets
on the dimer bonds are dominant excitations.
Several types of repulsive interactions between
these dimer triplet excitations can induce various insulating phases
and thereby yield magnetization plateaus.

\begin{figure}[tbp]
  \begin{center}
    \leavevmode
    \epsfig{file=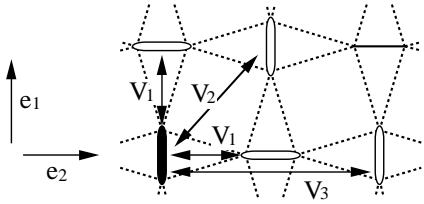,width=3.3in}
  \end{center}
  \caption{Two-body repulsive interactions up to the 3rd order of
    $J'/J$. 
$V_1=J'/2+J'^2/2J-J'^3/4J^2$, $V_2=J'^3/4J^2$,
    $V_3=J'^2/2J+3J'^3/4J^2$.}
  \label{fig:repulsion}
\end{figure}
For concreteness, we discuss
the Shastry-Sutherland model\cite{ShastryS}
shown in Fig.~\ref{fig:S-Slattice}.
The exact dimer ground state\cite{ShastryS} realizes for
$J'/J<0.69$.(Ref.~\onlinecite{MiyaharaUa}.)
Recently Kageyama et al.\cite{Kageyama} found this lattice
structure in SrCu$_2$(BO$_3$)$_2$ and
observed the magnetization plateaus
at $m/m_{\rm sat}=1/8$ and 1/4.   Because of the special
structure of the lattice, a triplet excitation is almost
localized.\cite{MiyaharaUa}
Considering the dimer triplet state with $S^z=1$ as a particle (a
hard-core boson by definition)
and the dimer singlet as a vacancy, we
derive an effective Hamiltonian for it using the perturbational
expansion from the $J'=0$ limit.
The expansion is performed up to the 3rd order in $J'/J$
from degenerate states with a constant number of dimer triplets.
The effective Hamiltonian up to the 2nd order is
\begin{eqnarray}
H &=& \biggl(J-B -\dfrac{J'^2}{J}\biggr)\sum_i n_i
+ \biggl(\dfrac{J'}{2}+\dfrac{J'^2}{2J}\biggr)
  \sum_{\langle i,j \rangle} n_i n_j \nonumber\\
&+& \dfrac{J'^2}{4J} \sum_{i\in A} \{ [b_i^\dagger(b_{i+e1}-b_{i-e1})+h.c.]
(n_{i-e2}-n_{i+e2}) \nonumber\\
& &+ 2 n_{i+e2}(1-n_{i})n_{i-e2}
+ (b_{i+e1}^\dagger b_{i-e1}+h.c.) n_{i} \} \nonumber\\
&+& \dfrac{J'^2}{4J} \sum_{i\in B} \{ \ \ \ e1 \leftrightarrow e2 \ \ \
\},
\end{eqnarray}
where $i(j)$ runs over an effective square lattice consisting of dimer
bonds (both horizontal and vertical) and horizontal (vertical) ones
belong to $A$ ($B$) sublattice.
The full form of the effective Hamiltonian up to the 3rd order will
be published elsewhere.\cite{MomoiT} The derived
Hamiltonian does not have the one-particle hopping
term (as was already reported in ref.~\onlinecite{MiyaharaUa}),
but contains many correlated-hopping processes,
where an effective hopping of a particle is mediated by another one.
This is one of our main observations.
Most 3rd-order terms concern the correlated hopping.
Longer-range repulsions between particles appear from
higher-order perturbations.
Diagonal repulsive interactions up to the 3rd order
in $J'/J$ are shown graphically in Fig.~\ref{fig:repulsion}.
The resulting Hamiltonian does not have $90^\circ$ rotational
invariance, since the lattice structure has low symmetry, and this may
lead to highly anisotropic CDW states.

\begin{figure}[tbp]
  \begin{center}
    \leavevmode
    \epsfig{file=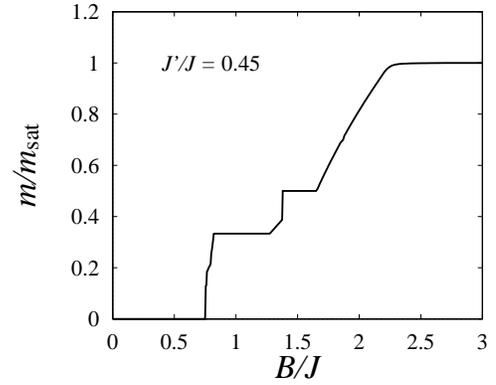,width=2.8in}
  \end{center}
  \caption{Magnetization process of the Shastry-Sutherland model with
    $J'/J=0.45$. }
  \label{fig:mag_pro_SS}
\end{figure}
\begin{figure}[btp]
  \begin{center}
    \leavevmode
    \epsfig{file=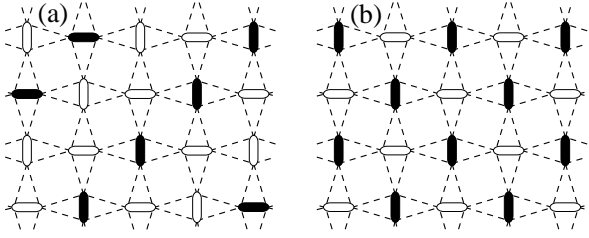,width=3.3in}
  \end{center}
  \caption{Spin configurations at magnetization plateaus at
    $m/m_{\rm sat}=1/3$ (a), 1/2 (b). Black bonds denote
    dimer triplet excitations.}
  \label{fig:CDW_SS}
\end{figure}

%%%%
We study the effective Hamiltonian in the classical limit.
To this end, we map the hard-core boson system to
the $S=1/2$
quantum spin system and then approximate the spin-1/2 by
a classical unit vector.
We search for the ground state with large sublattice structures
(e.g. a stripe-like one with 6-sublattice) both with
the mean-field approximation and a Monte Carlo method by decreasing
temperatures gradually.
The evaluated magnetization process is shown in
Fig.~\ref{fig:mag_pro_SS}.
Note a clear difference between the high- and low-field
region.
There appear plateau structures at
$m/m_{\rm sat}=1/2$ and 1/3. The plateau states have
CDW long-range orders shown in Fig.~\ref{fig:CDW_SS}.
Configurations realized for $m/m_{\rm sat}=1/2$ and 1/3
correspond to perfect closed packings provided that particles
avoid repulsion from 1st- and 2nd-order perturbation,
respectively.
The plateau at $m/m_{\rm sat}=1/2$
appears only in the region $0<J'/J<0.50$ and,
for large $J'/J$ the CDW is destroyed by the correlated hoppings,
which are dominant in the higher
order terms.  The correlated hoppings are so efficient
also at large particle density that
any plateau does not appear for $1/2<m/m_{\rm sat}<1$.
Below $m/m_{\rm sat}=1/3$, the correlated hoppings
occur rarely, because of a low particle density.
The observed 1/4-plateau (and 1/8-plateau)
of SrCu$_2$(BO$_3$)$_2$ may be formed
by weak longer-range repulsions
which are not taken into account in the
present study.
Recently Miyahara and Ueda discussed
semi-phenomenologically that the 1/4-plateau might come from a
CDW state with a stripe structure.\cite{MiyaharaUb}
In our approach, the repulsive
interaction relevant to the stripe CDW
may come from the higher-order terms
in perturbation, otherwise from other spin interactions that are
not considered in the Shastry-Sutherland model.
This remains to be a future problem.

\underline{\it Plaquette triplet:}
When the interactions or a special geometry of the lattice
allows four-spin plaquettes, in each of which four spins are
coupled more strongly than to the others,
individual plaquettes form singlets in the ground state.
%When four spins are coupled by strong interactions
%or they are strongly connected with each other due to geometry of the
%lattice, every four-spin plaquette forms a singlet in the ground
%state.
The triplet
states with $S^z=1$ on plaquettes are dominant excitations
in a weak magnetic field.
The insulator-conductor transition of these excitations can
take place thereby producing magnetization plateaus.

\begin{figure}[tbp]
  \begin{center}
    \leavevmode
    \epsfig{file=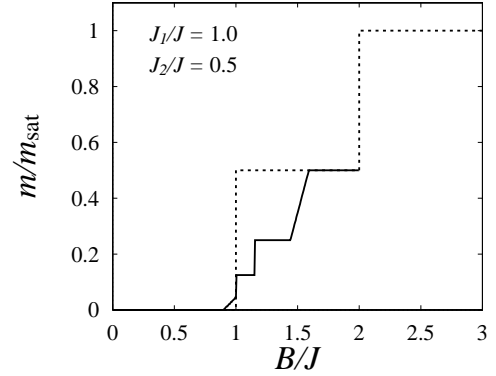,width=2.8in}
  \end{center}
  \caption{Magnetization process up to $m/m_{\rm sat}=0.5$ in the
    1/5-depleted square lattice with $J_1/J=1$ and $J_2/J=0.5$.
    The dotted line shows the case $J_1=J_2=0$.}
  \label{fig:mag_pro_1/5depl}
\end{figure}

An example of the plaquette singlet ground states is seen in
the $S=1/2$ HAF on the 1/5-depleted square lattice,\cite{SigristU}
which includes a possible
model for CaV$_4$O$_9$ as a special case.
The lattice is shown in Fig.~\ref{fig:1/5-depleted}.
In the isolated plaquette limit $J_1 = J_2 = 0$, a trivial plateau
already appears at $m/m_{\rm sat}=1/2$ for $J<B<2J$ (see
Fig.~\ref{fig:mag_pro_1/5depl}),
where every plaquette is in the triplet excited state with
$S^z=1$.\cite{FukumotoO}
When $m/m_{\rm sat}<1/2$, the triplet excitations (particle)
tend to hop if $J_1$ and $J_2$ are turned on,
and at specific (commensurate) values of $m/m_{\rm sat}$
they can show insulator-conductor transitions as a consequence of
the competition between the hopping and the repulsive
interaction.  Above $m/m_{\rm sat}=1/2$,
a plaquette quintuplet ($S=2$) with $S^z=2$ behaves as a particle and
can show magnetization plateaus between $1/2<m/m_{\rm sat}<1$.
In the following, we focus on a weak magnetic-field region, which
corresponds to the magnetization $0<m/m_{\rm sat}<1/2$.
Regarding the plaquette triplet excitation with $S^z=1$
as a particle
and the singlet state as a vacancy, we derive the effective
Hamiltonian of the particle by the 2nd-order perturbation
around the limit $J_1=J_2=0$.
%\begin{eqnarray}
%H &=& ( J_0 - \mu B) \sum_{ i } n_i \nonumber\\
%&+& \Biggl(\dfrac{J_1-2J_2}{6}+\dfrac{7J_1^2-8J_2^2}{144J_0}\Biggr)
%\sum_{\langle i,j \rangle} (b_i^\dagger b_j + h.c.)\nonumber\\
%&+& \Biggl(\dfrac{J_1+2J_2}{16}
%       -\dfrac{59 J_1^2-4700J_1J_2 + 5228J_2^2}{27648J_0}\Biggr)
%\sum_{\langle i,j \rangle} n_i n_j \nonumber\\
%&-& \dfrac{(J_1-2J_2)^2}{72J_0} \sum_{\langle i,j \rangle \atop \in N.N.N.}
%(b_i^\dagger b_j + h.c.)
%+ \dfrac{J_1^2}{48J_0} \sum_{\langle i,j \rangle \atop \in N.N.N.}
%n_i n_j \nonumber\\
%&+& \dfrac{J_1^2+4J_1J_2-4J_2^2}{216J_0}
%\sum_{\langle i,j \rangle \atop \in 3rd.N.N.}
%(b_i^\dagger b_j + h.c.)\nonumber\\
%&-& \dfrac{3J_1^2-2J_1J_2-2J_2^2}{192J_0}
%\sum_{\langle i,j \rangle \atop \in 3rd.N.N.} n_i n_j + others,
%\end{eqnarray}
%where ``$others$'' means correlated hopping and three-body interaction
%terms.
(The explicit form will be shown elsewhere.\cite{MomoiT})
The repulsive interactions range from a plaquette
to its nearest- and next-nearest neighbors.
If parameters satisfy $J_1 \simeq 2 J_2$, the hopping term is weak and
hence the system is in the strong coupling regime.
Then we may expect that the triplets crystallizes and
the magnetization plateaus appear at
$m/m_{\rm sat}=1/8$ and 1/4.
The mean-field approximation of the effective Hamiltonian indeed shows
magnetization plateaus at $m/m_{\rm sat}=1/8$ and 1/4 (see
Fig.~\ref{fig:mag_pro_1/5depl}).  They come from the insulating
phases with CDW long-range order of a square structure.
%%%%%%%%
A remark is in order here about the relevance of our results
to CaV$_{4}$O$_{9}$.   Quite recently, it was
shown\cite{Pickett}
that plaquettes of another type
({\em metaplaquettes}) consisting of $J_{2}$-bonds play
the main role in CaV$_{4}$O$_{9}$ contrary to earlier
studies\cite{SigristU}.
The physics is, however, almost the same also in this case;
the metaplaquette excitations behave like
particles and, $J$- and $J_{1}$-bonds 
induce hopping, and so on.
The detailed results will be reported in a longer paper.%
\cite{MomoiT}

\underline{\it Common features:}
To conclude this paper, we discuss a few features shared by the three
examples. According to an analogy to many-particle theories, a plateau
state corresponds to a CDW insulating state and gapless ones to
supersolids.
%Varying the parameters with magnetization
%fixed, we have a plateau-vanishing transition, where the off-diagonal
%superfluid order emerges. This kind of transition would be described
%by the (2+1)-dimensional classical XY model.
%
As the plateau state collapses by increasing the applied field,
superfluidity appears, whereas CDW exists in both phases.
%We next
%consider how the plateau state collapses into a supersolid or a
%superfluid by increasing the applied field.
Let us consider the case of 2nd
order transition, where the magnetization changes continuously.
Assuming that the on-set of superfluidity is well described by the
effective Hamiltonian of
the ordinary bosons with a short-range repulsion for low
energies, we conclude this transition is of the dynamical exponent
$z=2$;\cite{MFisherWGF} magnetization increases linearly like
$|H-H_{c}|$ apart from possible logarithmic corrections. (Note that the
form is quite different from that in 1D.)

{\bf Note added in proof:} K.\ Onizuka {\it et at.} recently observed a 
clear 1/3 plateau in SrCu$_2$(BO$_3$)$_2$, which we had predicted in 
this paper and had argued to be of a stripe structure.

\bigskip

We would like to thank the late Dr.\ Nobuyuki Katoh for stimulating
discussions at the beginning of this study. We also thank Hiroshi
Kageyama and Kenn Kubo for useful comments.


\begin{thebibliography}{99}

\bibitem[*]{address1} On leave of absence from Institute of Physics, 
   University of Tsukuba, Tsukuba, Ibaraki 305-8571, Japan.

\bibitem[**]{address2} Present address: Department of
   Physics, Kyushu University, Hakozaki, Higashi-ku, Fukuoka-shi,
   812-8581 Japan.

\bibitem{NarumiHSKNT} Y.\ Narumi, M.\ Hagiwara, R.\ Sato, K.\ Kindo,
  H.\ Nakano, and M.\ Takahashi, Physica B{\bf 246-247}, 509 (1998)
  and references cited therein.

\bibitem{Totsuka}
  K.\ Totsuka, Phys.\ Rev.\ B {\bf 57}, 3454 (1998).

\bibitem{NishimoriM} H.~Nishimori and S.~Miyashita, J.\ Phys.\ Soc.\
  Jpn.\ {\bf 55}, 4448 (1986).

\bibitem{Chubukov-Nikuni} A.V.\ Chubukov and D.I.\ Golosov, J.\ Phys.\
  Cond.\ Matter {\bf 3}, 69 (1991); A.\ E.\ Jacobs, T.\ Nikuni and H.\
  Shiba, J.\ Phys.\ Soc.\ Jpn.\ {\bf 62}, 4066 (1993).

\bibitem{KuboM}
  K.\ Kubo and T.\ Momoi, Z.\ Phys.\ B.\ {\bf 103}, 485 (1997);
  T. Momoi, H. Sakamoto, and K. Kubo, Phys. Rev. B, {\bf 59}, 9491
  (1999).

\bibitem{SuematsuOSSMD}
  H.\ Suematsu, K.\ Ohmatsu, K.\ Sugiyama, T.\ Sakakibara, M.\
  Motokawa and M.\ Date, Solid State Commun. {\bf 40}, 241 (1981);
  T.\ Sakakibara, K.\ Sugiyama, M.\ Date and H.\ Suematsu, Synthetic
  Metals, {\bf 6}, 165 (1983).

\bibitem{NojiriTM}
  H.\ Nojiri, Y.\ Tokunaga and M.\ Motokawa, J.\ Phys.\ (Paris) {\bf
    49} Suppl. C8, 1459 (1988).

\bibitem{Kageyama} H.\ Kageyama, K.\ Onizuka, Y.\ Ueda, N.V.\
  Mushnikov, T.\ Goto, K.\ Yoshimura, and K.\ Kosuge, Phys.\ Rev.\
  Lett.\ {\bf 82}, 3168 (1999).

\bibitem{MSE} See for example,
  M.\ Roger, C.\ B\"auerle, Yu.\ M.\ Bunkov, A.-S.\
  Chen, and H.\ Godfrin, Phys.\ Rev.\ Lett.\ {\bf 80}, 1308 (1998) and
  references cited therein.

\bibitem{ShastryS} S.\ Shastry and B.\ Sutherland, Physica {\bf 108}B,
  1069 (1981).

\bibitem{SigristU} N.\ Katoh and M.\ Imada, J.\ Phys.\ Soc.\ Jpn.\ {\bf
  64}, 4105 (1995); K.\ Ueda, H.\ Kontani, M.\ Sigrist and P.A.\ Lee,
  Phys.\ Rev.\ Lett.\ {\bf 76}, 1932 (1996).

\bibitem{MatsubaraM}
  T.\ Matsubara and H.\ Matsuda, Prog.\ Theor.\ Phys.\ {\bf 16}, 569
  (1956).

\bibitem{MiyaharaUa} S.\ Miyahara and K.\ Ueda,
  Phys.\ Rev.\ Lett.\ {\bf 82}, 3701 (1999).

\bibitem{MomoiT} T.\ Momoi and K.\ Totsuka, in preparation.

\bibitem{MiyaharaUb} S.\ Miyahara and K.\ Ueda, private communication.

\bibitem{FukumotoO}
  Y.\ Fukumoto and A.\ Oguchi, preprint.

\bibitem{Pickett} W.E.~Pickett, Phys.Rev.Lett. {\bf 79},
1746 (1997).

\bibitem{MFisherWGF} M.P.A.\ Fisher, P.B. Weichman, G.\ Grinstein, and
D.S.\ Fisher, Phys.\ Rev.\ B {\bf 40}, 546 (1989).

\bibitem{Onizuka} K.\ Onizuka et al., preprint.

\end{thebibliography}
\end{document}